\documentclass{aa}  
\usepackage{graphicx}
\usepackage{textcomp}

\begin{document}

\title{Dusty globules in the Crab Nebula
          \thanks{Based on observations collected with the NASA/ESA Hubble Space Telescope, obtained at the Space Telescope Science Institute.} }
         

   \author{T. Grenman\inst{1}
           \and
           G. F. Gahm\inst{2}
           \and
           E. Elfgren\inst{1}
           }


   \institute{Division of Physics, Lule\aa\ University of Technology, SE-97187 Lule{\aa}, Sweden\\
              email: \mbox{tiia.grenman@gmail.com}
              \and
             Stockholm Observatory, AlbaNova University Centre, Stockholm University,
              SE-106 91 Stockholm, Sweden   }

   \date{}

   

 \abstract
   {Dust grains are widespread in the Crab Nebula. A number of small dust concentrations are visible as dark spots against the background of continuous synchrotron emission in optical images.} 
   {Our aim is to catalogue such roundish, dusty globules and investigate their properties. }
   {From existing broad-band images obtained with the Hubble Space Telescope, we located 92 globules, for which we derived positions, dimensions, orientations, extinctions, masses, proper motions, and their distributions. }
   {The globules have mean radii ranging from 400 to 2000 AU and are not resolved in current infrared images of the nebula. The extinction law for dust grains in these globules matches a normal interstellar extinction law. Derived masses of dust range from 1 to $60 \cdot 10^{-6}$ $M_\odot$, and the total mass contained in globules constitute a fraction of approximately 2 \% or less of the total dust content of the nebula. The globules are spread over the outer part of the nebula, and a fraction of them coincide in position with emission filaments, where we find elongated globules that are aligned with these filaments. Only 10 \% of the globules are coincident in position with the numerous H$_2$-emitting knots found in previous studies.   All globules move outwards from the centre with transversal velocities of 60 to 1600 km s$^{-1}$ , along with the general expansion of the remnant. We discuss various hypotheses for the formation of globules in the Crab Nebula.}


 \maketitle
%
\section{Introduction}
\label{sec:intro}

Dust ejected from supernovae has been considered to play an important part in supplying the interstellar medium with dust grains (e.g., Kozasa et al. \cite{koz09}). Substantial amounts of dust have indeed been found in young supernova remnants based on observations of infrared emission from dust grains. For Cas A (e.g., Roh et al. \cite{rho08}) and SN 2003gd  (e.g., Sugerman et al. \cite{sug06}) the total amount of dust was estimated  to be up to a few times 10$^{-2}$ $M_\odot$. For SNR 1987A, Ercolano et al. (\cite{erc07}) obtained an upper limit for dust mass of $7.5\cdot  10^{-4}$ $M_\odot$, assuming a clumpy distribution of dust, at a phase two years after the explosion. Indebetouw et al. (\cite{ind14}) and Matsuura et al. (\cite{mat15}), however, found much larger amounts of dust, of the order of 0.5 $M_\odot$, from observations obtained at later phases. It was proposed that the dust is formed in the inner ejecta, the cold remnants of the exploded stellar core, and that grain growth occur in the remnant at later phases.

Here, we will focus on another young and well-known dust-containing supernova remnant named the Crab Nebula, formed after the supernova in 1054 AD. Based on arguments concerning the low expansion velocities of the ejecta and the lack of signatures of remote, fast-moving gas (Lundqvist et al. \cite{lun86}, \cite{lun12}; Wang et al. \cite{wan13}) in addition to the energy budget of the complex (e.g., Yang \& Chevalier \cite{yan15}), it is has been proposed that the progenitor was a star in the more massive part of the asymptotic giant branch (Moriya et al. \cite{mor14}). At the centre of the nebula lies one of the first known pulsars, PSR B0531+21, surrounded by an expanding synchrotron radiating nebula and a network of filaments radiating in emission lines from various ions. The general properties of the Crab Nebula have been reviewed by Davidson \& Fesen (\cite{dav85}) and Hester (\cite{hes08}). 

The velocity field over the expanding remnant has been mapped from proper motions of the optical filaments (Trimble 1968; Nugent \cite{nug98}) and related radial velocities (Wykoff et al. \cite{wyk77}). Bietenholz \& Nugent (\cite{bie15}) also  included radio measurements to conclude that the synchrotron nebula, fed by energy from a pulsar-wind nebula (see B\"uhler \& Blandford \cite{buh14}), has been strongly accelerated compared to the filaments.   

The presence of dust in the Crab Nebula was first inferred from the infrared (IR) excess emission over the synchrotron spectrum noted in Trimble (\cite{tri77}). With the launches of the Spitzer Space Telescope and the Herschel Space Observatory, the distribution of IR-emitting dust could be mapped in great detail (e.g., Temim et al. \cite{tem06}, \cite{tem12}; Temim \& Dwek \cite{tem13}; Gomez et al. \cite{gom12}; Owen \& Barlow \cite{owe15}). Dust is spread over the entire nebula, and strongly concentrated to the optical filaments. The total dust content was estimated by these authors to be in the order of a tenth of a solar mass. Sankrit et al. (\cite{san98}) proposed that dust is freshly formed in filaments resulting from Rayleigh-Taylor instabilities acting on the ejecta.

Dust was also recognized in optical images as dark features silhouetted against the bright background in Woltjer \& V\'eron-Cetty (\cite{wol87}). From an examination of photographic plates, Fesen \& Blair (\cite{fes90}) catalogued 24 ``dark spots'' and measured the amount of extinction in them, and Hester et al. (\cite{hes90}) draw attention to several optically visible dusty blobs and filaments. One such blob attached to a dusty filament, coincident with an optically bright filament, was investigated in Sankrit et al. (\cite{san98}) from images taken with the Hubble Space Telescope (HST). Several extended irregular dusty filaments can be seen in optical continuum HST images and also in narrow band images covering certain emission lines (Blair et al. \cite{bla97}). In addition to these dusty structures, Loh et al. (\cite{loh11}) found a number of knots emitting in the line of molecular hydrogen at 2.12 $\mu$m, further investigated in Richardson et al. (\cite{ric13}). Barlow et al. (\cite{bar13}) also mapped emission  from the molecule ArH$^{+}$ from several positions in the nebula.
 
In the present study we focus on the roundish, dusty cloudlets seen in optical images of the Crab Nebula. These are reminiscent of the so-called globulettes present in \ion{H}{ii} regions surrounding young stellar clusters as studied in  Gahm et al. (\cite{gah07}) and Grenman \& Gahm (\cite{gre14}), for example, who concluded that the globulettes are remnants of eroding molecular shells and pillars surrounding the clusters. The Crab Nebula objects have a different origin, and may differ in chemical composition, and in the following we refer to them as {\it globules}. We have located 92 globules in the nebula from continuum images taken with HST. Our aim has been to define central positions, average radii, and the extinction due to dust, providing a measure of the associated extinction law. From these data we have estimated the mass of dust contained in the globules. In addition, we have measured the proper motion of the globules and determined their transversal velocities from images taken between 1994 and 2014. 

The paper is organized as follows. We present the fields containing globules and their measured properties in Section~\ref{sec:obs}. In Section~\ref{sec:results} we derive extinction and masses, and discuss the distribution and motion of the globules, as discussed further in Section~\ref{sec:disc}. We end with a summary in Section~\ref{sec:conclude}.

\section{Archival HST images and measurements}
\label{sec:obs}

\begin{figure}[t] 
\centering
\resizebox{8cm}{!}{\includegraphics[angle=00]{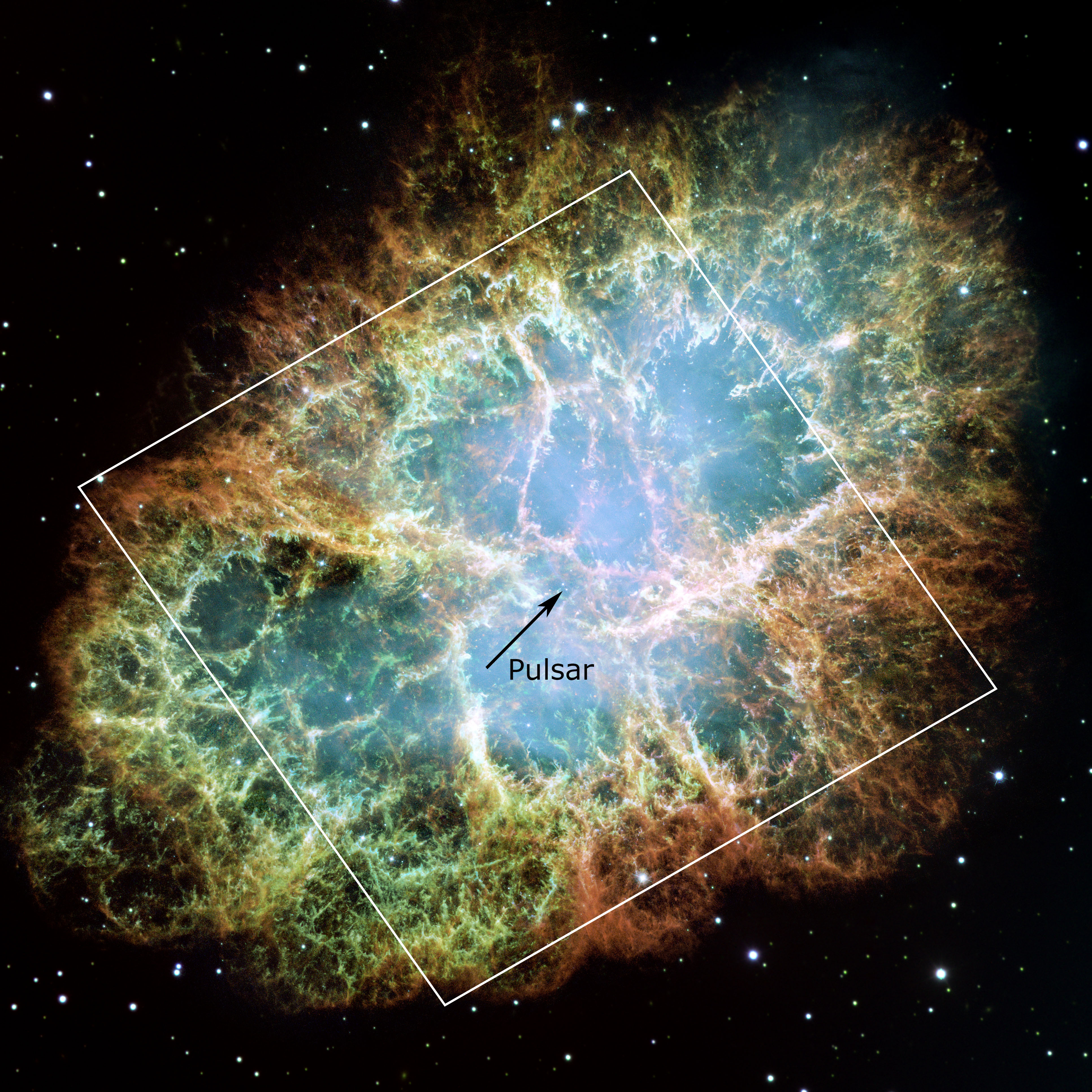}}
\caption{HST color image composed from images in the F502N, F631N, and F673N filters (courtesy: NASA, ESA, and J. Hester, Arizona State University). The area investigated here is marked, and the arrow points to the location of the pulsar (north is up and east left).} 
\label{crab}
\end{figure} 

\begin{figure}[t] 
\centering
\resizebox{9.2cm}{!}{\includegraphics[angle=00]{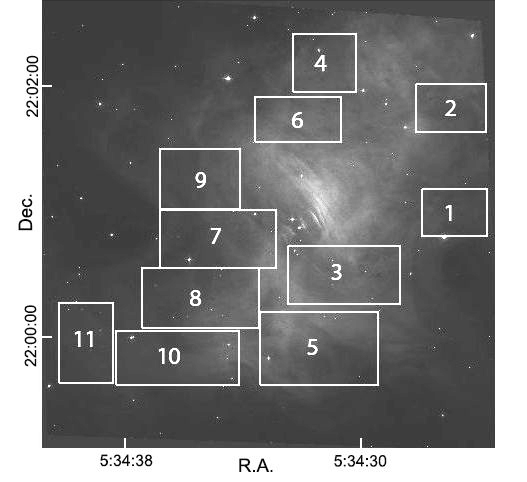}}
\caption{The rectangles on this continuum image mark the fields in which globules were found. The positions and dimensions of each box are listed in Table~\ref{tab:regions}. } 
\label{fields}
\end{figure} 

\begin{figure}[t] 
\centering
\resizebox{8cm}{!}{\includegraphics[angle=00]{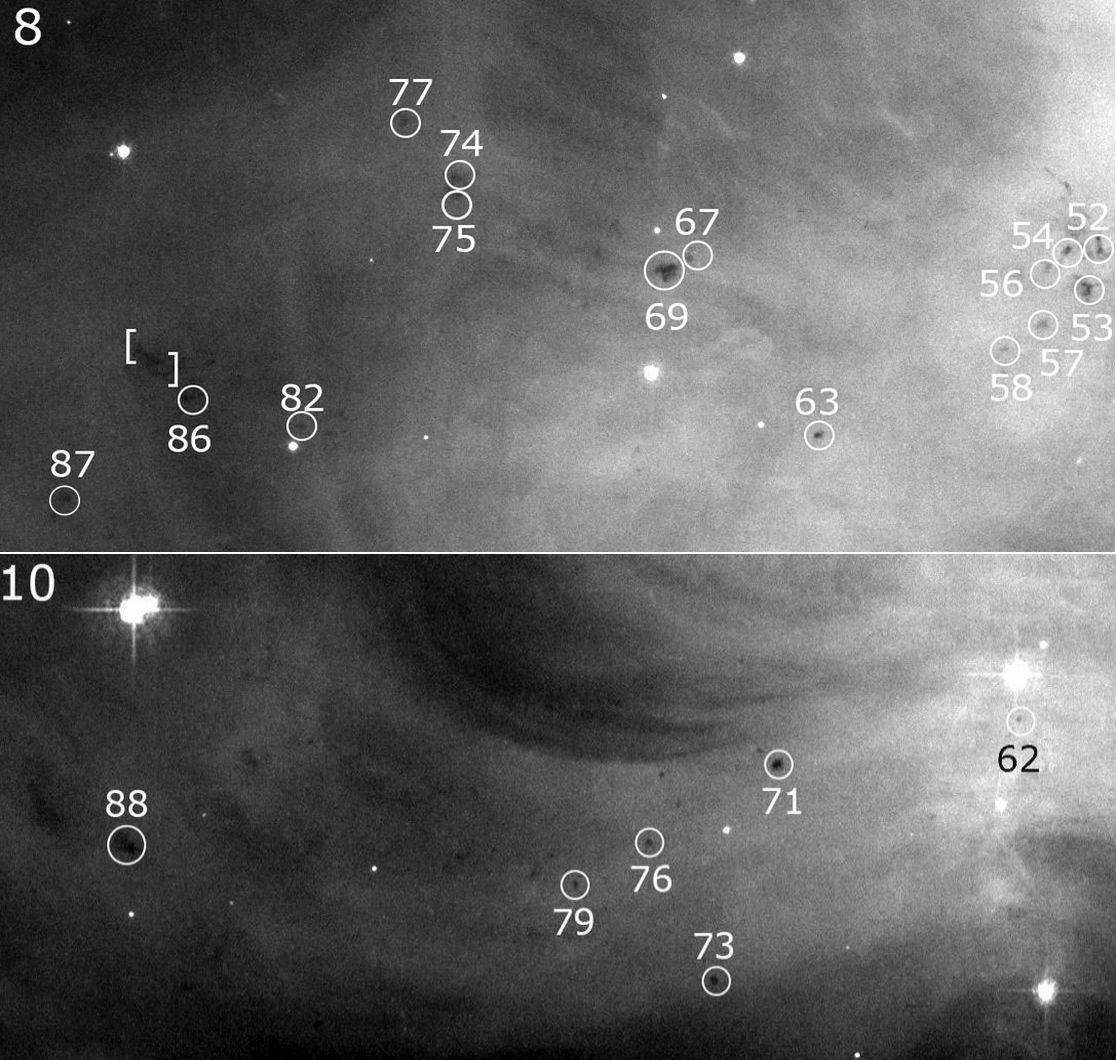}}
\caption{Close-ups of fields 8 and 10 (marked in Fig.~\ref{fields}). Globules are marked with circles and CrN number. The brackets enclose some irregular, filamentary features. } 
\label{smallFields}
\end{figure} 

\begin{figure}[t] 
\centering
\resizebox{8cm}{!}{\includegraphics[angle=00]{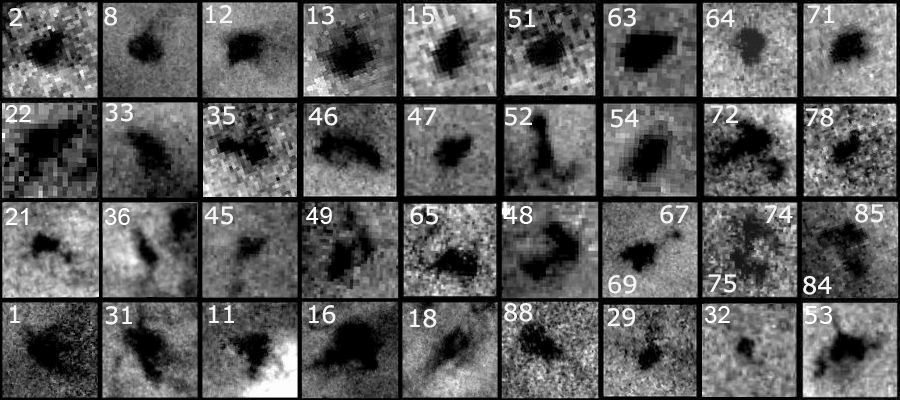}}
\caption{Examples of globules. The image size differs from panel to panel but spans 1\arcsec - 4\arcsec. } 
\label{mosaic}
\end{figure} 

\begin{table*}
\centering
\caption{Archival HST /HLA  data used.}
\begin{tabular} {lcccc }
 \hline
     \noalign{\smallskip}
      Program ID & Filter & Exposure time (s) & Year &Archive Data Set \\       \noalign{\smallskip}
\hline
         \noalign{\smallskip}
5206 & F547M&1000 &1994& u24r0404t\\
5354 &  F547M&800 - 1000 &1995&u2bx0501t - u2bx0504t     \\
6129 &  F547M& 1000&1995 - 1996&u2u60101t - u2u60702t    \\
         & F814W &      700, 1200&1995 - 1996   &u2u60106t, u2u60504t\\
7407  &F547M&1200&2000 - 2001&u50v0302r, u50v1201r, u50v2101r, u50v5902m\\
8222 &F547M&900&2000&u5d10201r - u5d10601r\\
13510&  F550M&2000 &2014 &jcgt03010\\
13772&  F550M&2000 &2014 &jcp301010 \\
        \noalign{\smallskip}
\hline
\end{tabular}
\label{tab:data} 
\end{table*} 

\begin{table}

\centering
\caption{Central positions and sizes of the fields containing globules.}
\begin{tabular} {lcc }
\hline
     \noalign{\smallskip}
  Field & Central Position &Size  \\
  & R.A. (2000)  Dec (2000)&(arcsec)\\
\hline
     \noalign{\smallskip}
1&5:34:26.72 +22:00:58.6& 22x30\\
2&5:34:26.89 +22:01:47.1&28x34\\
3&5:34:30.42 +22:00:29.4&52x27\\
4&5:34:31.05  +22:02:08.0&28x28\\
5&5:34:31.24  +21:59:55.4&34x55\\
6&5:34:31.95  +22:01:41.8&21x40\\
7&5:34:34.63  +22:00:46.4&27x54\\
8&5:34:35.21  +22:00:18.6&28x54\\
9&5:34:35.24 +22:01:14.1&28x37\\
10&5:34:35.98   +21:59:50.8&25x57\\
11&5:34:39.04   +21:59:57.9&37x25\\
     \noalign{\smallskip}
\hline
\end{tabular}
\label{tab:regions} 
\end{table} 

Optical images of the Crab Nebula were downloaded from the HST archive and from the Hubble Legacy Archive (HLA). Observations from programs GO-5206, 5354, 6129, 7407, and 8222 were obtained with the WFPC2 camera and from GO-13510 and 13772 with the ACS/WFC camera. The pixel sizes correspond to approximately 0.05 arcsec pixel$^{-1}$ and 0.1 arcsec pixel$^{-1}$ for the ACS/WFI and WFPC2 camera, respectively. For further details about the image reductions made at the HST centre, see the HST Data Handbooks (https://archive.stsci.edu/hst/). 

Areas containing dusty globules were selected for further analysis. In particular, we inspected images in filters not contaminated by strong emission lines, namely the F547M, F550M, and F814W filters. Comparisons were also made with images covering various emission lines. Table~\ref{tab:data} summarizes the observational material used, including dates, exposures, and image designations according to the HST/HLA archive. The exposures date from 1994 to 2014, hence covering a timespan of 20 years. The fields in which we found globules are marked in Fig.~\ref{fields}, and these are listed with central positions and sizes in Table~\ref{tab:regions}. Each globule is designated as CrN plus a number. Close-ups of all fields are shown in Appendix~B, where globules are numbered and marked with circles, and two examples, fields 8 and 10, are shown in Fig.~\ref{smallFields}. Fig.~\ref{mosaic} shows examples of individual globules. As a rule, they appear as distinct dark patches against the bright synchrotron continuum. Most of the globules are roundish and slightly elliptical, but some objects have more irregular shapes. We also draw attention  to a smaller number of tiny, dark, thin filaments, and a few diffuse, patchy structures marked with clammers in the figures, but not included in the subsequent analysis.

\begin{figure}[t] 
\centering
\resizebox{8cm}{!}{\includegraphics[angle=00]{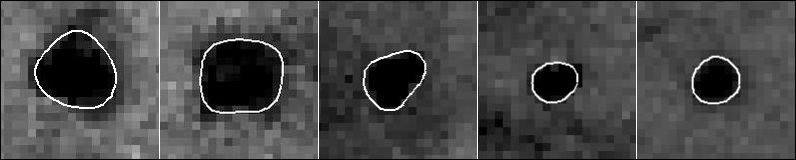}}
\caption{Examples of globules, where the white contour defines the shape of each object. } 
\label{map}
\end{figure} 

Proper motions of globules were derived from images collected over the entire period, spanning a total of 20~years. The accuracy of these measures was estimated to $\sim~0.045\arcsec$~yr$^{-1}$ from the x and y pixel offsets of reference stars in both the first and last epoch images. The transversal velocity (in km s$^{-1}$) is $V_t = 4.74\cdot\mu\cdot D$, where $\mu$ is the proper motion (in arcseconds/year) and $D$ is the distance (in parsecs), which has been been estimated to $\sim$ 2 kpc (e.g., Trimble~\cite{tri68}; Davidson \& Fesen \cite{dav85}).

\begin{table*}
\centering
\caption{Properties of the globules in the Crab Nebula$^a$.}
\begin{tabular} {lccccccccccc}
\hline
     \noalign{\smallskip}
CrN & Field & R.A. & Dec. & $\bar r$ & $\bar R$ & P.A. &  $\mu_{\alpha}$ & $\mu_{\delta}$ & V$_t$ & $M_{d}$ &Remarks\\
 && J(2000.0) &J(2000.0) & ($\arcsec$) &  (AU) & (degr.)& ($\arcsec$/yr)&($\arcsec$/yr) & (km/s) & ($M_\odot$$\cdot 10^{-6}$) &\\
\hline
\noalign{\smallskip} 
  
1 &2&   5:34:25.77 &+22:01:40.1 &       0.5&    1000&   41  &           -0.163  &       +0.064  &1660 &       10 &     C\\
2 &1    &5:34:25.86&+22:01:03.6 &       0.3&     600&   -16     &  -0.133       &       +0.016  &1270&      3  &               C\\
3 &     2&5:34:26.25&   +22:01:56.1     &       0.3&     600&           &                -0.145  &       +0.077  & 1522& 3       &               \\
4 &2    &5:34:26.31&+22:01:54.9 &       0.2&     400&           &       -0.134  &       +0.093  & 1546&   1       &       \\
5&      2&5:34:26.34&   +22:01:54.3     &       0.2&     400&   -33     &                -0.143  &       +0.078  & 1544& 2       &       \\
6&1&    5:34:26.84&     +22:00:58.3     &       0.2&     400&           &                -0.113  &       +0.009  & 1075& 2       &       C  \\
7&      1&5:34:26.89&   +22:01:00.4     &       0.3&     600&   -42     &               -0.114  &       +0.016  & 1091  & 3       &                             \\
8&2     &5:34:27.06&    +22:01:52.5     &       0.5&    1000&           &       -0.114  &       +0.071  & 1273&   12      &  Knot 52$^{1}$, 4E$^{2}$  \\           
9&      1&5:34:27.20&   +22:00:58.5     &       0.2&     400&           &       -0.113  &       +0.009  & 1075& 1  &      C to 10                \\
10&1&   5:34:27.21&     +22:00:57.7     &       0.4&     800&           &               -0.118  &       +0.004  & 1119 &  6& C to 9, Irr      \\
& & .......& & & & & & & & \\

\noalign{\smallskip}
\hline

\end{tabular}
\label{tab:first} 
\tablefoot{
\tablefoottext{a}{Continued in Appendix~A,}
\tablefoottext{1}{from Loh et al. (\cite{loh11}),}
\tablefoottext{2}{from Fesen \& Blair (\cite{fes90})}      
}
\end{table*} 

World system coordinates (WSCs) were obtained  for the WPFC2 and ASC images by using the Sloan Digital Sky Survey and 2MASS catalogs. The accuracy of the positions is $\sim$ 0.15\arcsec, and the typical absolute astrometric accuracy from the HLA images is $\sim 0.3\arcsec$ in each coordinate. In order to improve the accuracy of the coordinates of a given object in an ASC image, we convert the (x, y) positions in pixel space to equatorial coordinates by using an astrometric position calculator (Jordan D. Marche, http://www.phys.vt.edu/~jhs/SIP/astrometrycalc.html). In this program we entered positions of ten reference stars and the resulting rms residual  is 0.09\arcsec.

Central positions of globules were derived from images taken in 1994, 1995, 2000, and 2014. In one 2014 ACS image (GO-13772) there are several small objects, with radii less than 0.15'', which were not detected in the WFPC2 images. Some of these may be artifacts, and we decided to measure only objects with average diameters of $\geq$ 3 pixels. We define a shape from an outer contour set at a level where the intensity has dropped to 95$\%$ of the interpolated nebular background (see examples in Fig.~\ref{map}). Outside this contour the level of noise starts to affect the definition of the boundary, but, as a rule, very little matter resides in these outskirts. Ellipses were fitted to these contours from which we derive radii of the semi-major and semi-minor axes as a measure of the shape and size of a globule by using the method described in Grenman (\cite{gre06}), and the positions are defined as the centres of these ellipses. From the mean of these radii we assign a mean radius $\bar r$ based on measurements from two different years. For irregular objects we measured radii from long and short axes. Position angles (P.A.) were measured  in degrees, from north to east, but only for objects with ratios of major and minor axes $\geq$~1.5.   

Table~\ref{tab:first} shows the first ten entries of the complete table, Table A.1 in the Appendix. The field where the globules reside is shown in the second column. The following four columns give measured positions as obtained for the year 2014 followed by mean radii, $\bar r,$ expressed in arcseconds and $\bar R$ in AU (adopting a distance of 2 kpc). Column 7 gives position angle (P.~A.) for elongated objects. Columns 8, 9, and 10 give proper motions, $\mu_{\alpha}$ and $\mu_{\delta}$, and the associated transversal velocity V$_t$. Our estimates of the mass of dust in the globules are described in Sect.~\ref{sec:dustmass} and are given in column 11. The last column contains remarks, where $C$ indicates that the object is connected to another globule (numbered) by thin dusty threads, and $Irr$ marks an object of pronounced irregular shape. Objects that have been studied earlier by Fesen \& Blair (\cite{fes90}) are introduced by the number assigned by them. Loh et al. (\cite{loh11}) list a number of H$_{2}$-emitting knots, assigned $Knot$ plus a number, and five of these coincide in position with globules CrN 8, 12, 16, 30, and 31 as noted in the remarks. Globule CrN 8, identical with Knot 52 and 4E, forms the head of a dusty filament studied by Sankrit et al. (\cite{san98}). Also CrN 12, identical with Knot 51 and 4D, is part of a filamentary structure, which was studied in depth by Richardson et al. (\cite{ric13}).

\section{Results}
\label{sec:results}
           
\subsection{Extinction}
\label{sec:extinction}

\begin{table}
\centering
\caption{Extinction measures at 8140 \AA\ and 5470 \AA\ and ratios.}
\begin{tabular} {lccc }
\hline
     \noalign{\smallskip}
  CrN & $A_{8140}$ &  $A_{5470}$ & $A_{5470}$/$A_{8140}$ \\  
\hline
     \noalign{\smallskip}
     
8 & 0.144 & 0.233 & 1.62 \\
43 & 0.106 & 0.204 & 1.92 \\
47 & 0.148 & 0.277 & 1.87 \\
53 & 0.131 & 0.256 & 1.95 \\
63  & 0.156 & 0.285 & 1.83\\
71 & 0.173 & 0.321 & 1.86 \\
72 & 0.173 & 0.340 & 1.97\\

     \noalign{\smallskip}
\hline
\end{tabular}
\label{tab:extinction} 
\end{table} 

Images obtained with the F547M and F814W filters offer an opportunity to estimate the extinction law in the globules. First we measure the residual intensity for each pixel within a globule, $I$, relative to the interpolated bright background $I_{0}$ in the images. This value relates to the extinction due to dust at wavelength, $\lambda$, through $A_{\lambda} = -2.5\cdot \log(I/I_{0})$, provided that the globule lies in the foreground of the emission nebula. From the average value of $I$ we derive an average value of the extinction for individual globules at the central wavelengths of the filters, 5470 \AA\ and 8140 \AA.

Seven rather dark and distinct globules were selected for extinction measures, since these are likely to be located essentially in the foreground of the emission nebulosity. The resulting values are given in Table~\ref{tab:extinction} for the two wavelengths (columns 2 and 3), and their
ratios are listed in the fourth column. The mean value of these ratios is 1.86, which is close to 1.89,  the ratio expected from a standard interstellar reddening law (Savage \& Mathis \cite{sav79}). The spread around the mean is expected from uncertainties in measurements of average extinctions. Values $>$~1.9 do not fit any extinction law. Values $<$ 1.8 could indicate that the grains are larger than normal interstellar grains, but it could also mean that the globule is embedded at some depth inside the emitting bubble.  In this case, the extinction values must be calculated for a reduced continuum, since the globule is filled with foreground emission. 

With the exception of CrN 8, the extinction is in close agreement with that of a normal extinction law, which confirms our suspicion that the selected objects are not deeply embedded in nebulosity. Finally, we verified some of the darkest filaments using the same method applied to smaller areas within the filaments and found extinction ratios similar to those of the globules. We conclude that the extinction of dust grains in the globules is similar to that of normal interstellar grains.  

\begin{table}
\centering
\caption{Globule masses based on two different methods.}
\begin{tabular} {lcc}
\hline
     \noalign{\smallskip}
CrN & $M_{d}$$\arcmin$ & $M_{d}$(Table A)    \\
   & ($M_\odot$$\cdot$$10^{-6}$) & ($M_\odot$$\cdot 10^{-6}$)  \\
      \noalign{\smallskip}
\hline
        \noalign{\smallskip}
8     &  6  &    12   \\
12   &  9 &     16  \\
47      &    2  &  4  \\
63      &   2    &  3  \\
     \noalign{\smallskip}
\hline

\end{tabular}
\label{tab:mass} 
\end{table} 

\subsection{Dust masses}
\label{sec:dustmass}

In order to estimate the mass of dust in the globules we consider two extreme cases. In the first case we assume that the residual intensity in the darkest parts of a globule is due entirely to foreground emission. In the second case the object is in the foreground and residual intensities in the darkest areas trace only background emission. In both cases, we set the level of residual nebular emission to 95\% of the darkest area in the object in order to avoid infinite extinction in individual pixels due to noise. We then derive the extinction $A_{5470}$ for the two cases, and since the difference between the extinction in V and that derived from F547M images is negligible, $A_{5470}$ is a measure of $A_{V}$.  

The column density of dust in any sightline through a globule can be calculated once the properties of the grains are known. However, since the extinction through the globules appears to match a standard interstellar reddening law (Sect.~\ref{sec:extinction}) we will first follow an alternative method for deriving column densities of dust that was used for similar studies of globulettes in \ion{H}{ii} regions in Gahm et al. (\cite{gah07}) and Grenman \& Gahm (\cite{gre14}). This method is based on the relationship between column density of molecular hydrogen and $A_{V}$, namely $N(H_{2})$ = $9.40\cdot 10^{20} A_{V}$, found by Bohlin et al. (\cite{boh78}) for normal interstellar conditions, and in our case we apply a normal mass ratio of  dust to gas of 1/100 to obtain the dust masses. Assuming solar chemical composition, the total column density of dust expressed in g~cm$^{-2}$ mag$^{-1}$ amounts to $N _{d}$ = $4.3\cdot 10^{-7} A_{V }$.

We calculate the total column mass densities of dust for each pixel in individual globules and sum over all pixels inside the contours, as defined in Sect.~\ref{sec:obs}, to obtain the total mass of dust in each globule. Since we do not know how deeply the globules are embedded in the nebula we present the mean mass, $M_{d}$ , of the two extremes described above in column 11 in Table~\ref{tab:first}, and continued in Table A.1 in the Appendix. The difference in mass compared to the extremes is relatively small; a factor 2.5 on average. The dust masses range from one to several tens of earth masses. Summing over all objects we obtain a total dust mass contained in the Crab Nebula globules of $4.5\cdot 10^{-4}$ $M_\odot$. The total mass of individual globules depends on the unknown gas-to-dust ratio in these objects. Owen \& Barlow (\cite{owe15}) estimated a gas-to-dust ratio of approximately 30 from their studies of widespread dust. With this ratio the total mass of all globules in our study is $1.4\cdot 10^{-2}$ $M_\odot$.

From the mass ($M_{d}$) and volume of individual globules, we derive the mean densities of dust, $\bar {\rho}$, for all objects in Table~\ref{tab:first}. The result is that the densities range $8\cdot 10^{-22}$ $<$ $\bar {\rho}$ $<$ $7\cdot 10^{-21}$ g cm$^{-3}$. 
 
Next, we compare these results with calculations based on assumptions of grain properties. These relate to the extinction in magnitudes and grain properties through $A _{V}$~=~1.086 $N_{d }\cdot Q_{e }\sigma$, where $N _{d }$ is the column number density of grains, $Q _{e}$ is the extinction efficiency factor (Spitzer \cite{spi78}), and ${\sigma}$ is the cross section of a grain. We select four spherical objects with smooth density distributions and apply a grain radius of $2\cdot 10^{-5}$ cm. The corresponding cross section of a grain is then ${\sigma}$ = $1.3\cdot 10^{-9}$ cm$^{2}$. For the grain density we adopt a value of 3 g cm$^{-3}$ approximately corresponding to a mixture of the graphite and silicate materials used in Owen \& Barlow (\cite{owe15}) resulting in a grain mass of $m _{d}$ = $1.0\cdot 10^{-13}$ g. The extinction efficiency factor is set to $Q _{e} = 2$ and the column density of dust is $m_{c}$ = $N _{d}\cdot m _{d}$ (g~cm$^{-2})$. 

We find a mean value of $A _{V}$ from the intensity profiles across each globule in the F547 filter images, with the same assumptions on the amount of foreground emission as used above in the Bohlin relation\textless. The corresponding value of $N _{d}$ represents a mean value. From the column density $m_{c}$ , integrated over the surface area of the globule, we estimate the total mass of dust, $M_{d}$$\arcmin$, as listed in the second column of Table~\ref{tab:mass}, which can be compared with the values taken from Table~\ref{tab:first} (third column). The masses based on grain properties are approximately a factor two smaller than those derived from the Bohlin relation. Given the uncertainties in these assumptions, we find that the agreement is fair.

The masses are very sensitive to the choice of grain radius and, to some extent, to the grain density. Similar calculations were performed by Fesen \& Blair (\cite{fes90}), Sankrit et al. (\cite{san98}), and Loll et al. (\cite{lol10}) based on a smaller grain radius and a slightly lower grain density than that used here. Fesen \& Blair (\cite{fes90}) noted that the resulting masses become smaller than those inferred from infrared measurements, and also proposed that the grains are larger than their initial assumptions. The radii adopted here are somewhat larger than those assumed by Temim et al. (\cite{tem06}) to match the IR emission at 3.6 and 4.5 ${\mu}$m but in the lower range of radii assumed for dust produced in AGB winds (e.g., H\"offner et al. \cite{hof16}).

\subsection{Distributions of globules}
\label{distribution}

\begin{figure*}[t] 
\centering
\resizebox{9cm}{!}{\includegraphics[angle=00]{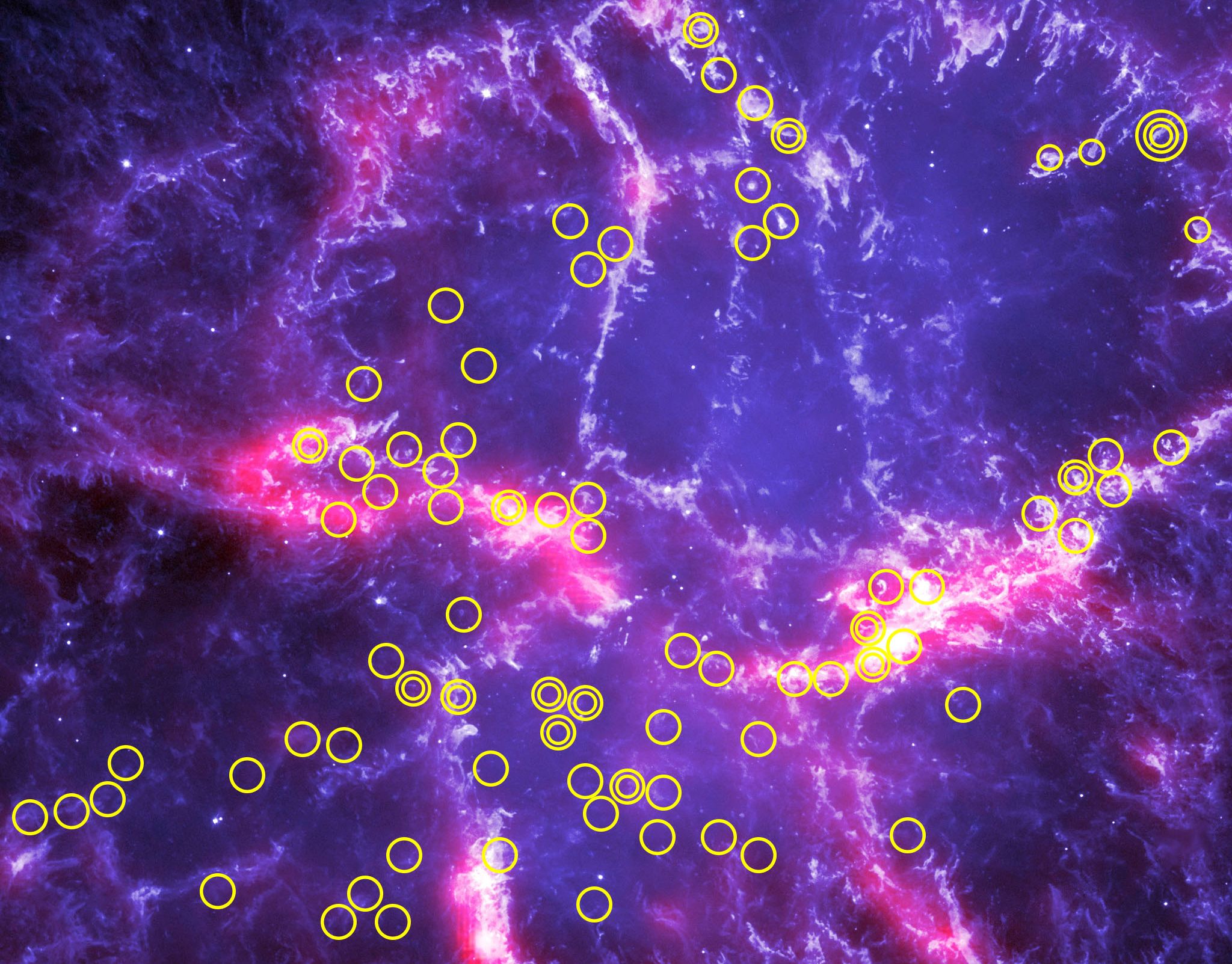}}
\resizebox{8.1cm}{!}{\includegraphics[angle=00]{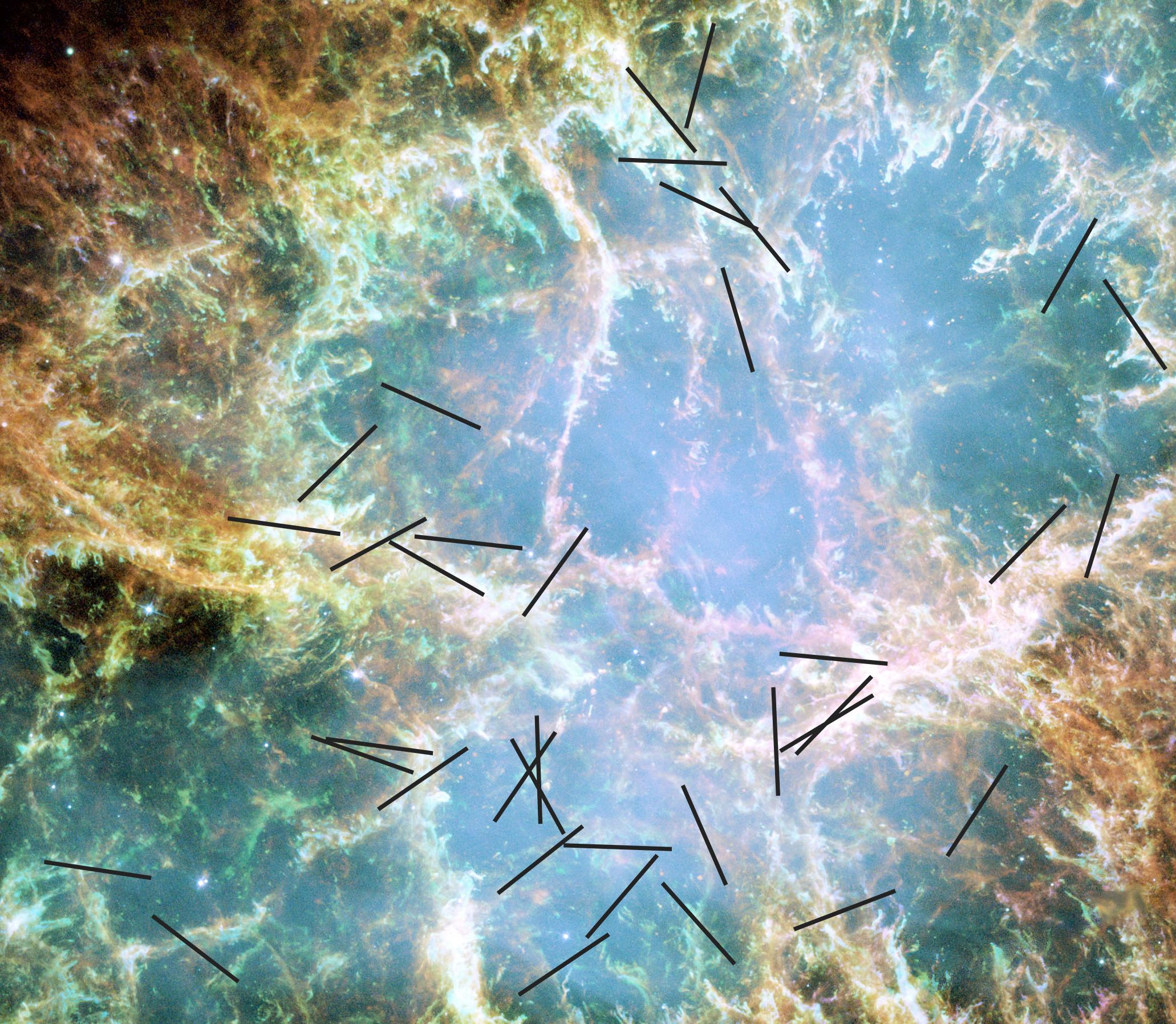}}
\caption{Left: Individual globules are marked with circles on this HST optical emission line image (blue-white) combined with a Herschel Space Observatory 70 $\mu$m image (red) -- credit: HST and Herschel Space Observatory, ESA, NASA. Right: Orientations of elongated objects on the background of the HST image shown in Fig.~\ref{crab}. } 
\label{fig:positions}
\end{figure*} 

We have catalogued a total of 92 globules. The majority of the objects have radii $<$1400 AU and the size distribution peaks at 500 AU. The dust masses range from 1~to $60\cdot 10^{-6}$~$M_\odot$, but very few objects have dust masses $>$~10$^{-5}$~$M_\odot$. Their distribution across the Crab Nebula is shown in the left panel of Fig.~\ref{fig:positions} on the combined HST and Herschel Space Observatory image, featuring the location of knots and filaments visible both in the optical emission lines and in the far infrared emission from dust. 

The globules are spread over the nebula, but the area surrounding the pulsar is void of globules, and the northern part of the nebula contains less objects than the southern part. Many globules fall along the nebular filaments, but a substantial fraction reside in areas outside the filaments. We note that among the 55 knots emitting H$_{2}$ found in Loh et al. (\cite{loh11}), only five  coincide with globules, and Richardson et al. (\cite{ric13}) considered cases where dust-free knots could be sources of strong H$_{2}$ emission.  

Position angles were measured for elongated objects (40 \% of the total sample), and the corresponding distribution over the nebula is shown in the right panel of Fig.~\ref{fig:positions}. The orientations are randomly distributed relative to the centre of the nebula. However, there are several examples of globules that appear to be aligned with certain emission filaments to the north and to the extended feature crossing the image just south of the centre.

\begin{figure}[t] 
\centering
\resizebox{8.5cm}{!}{\includegraphics[angle=00]{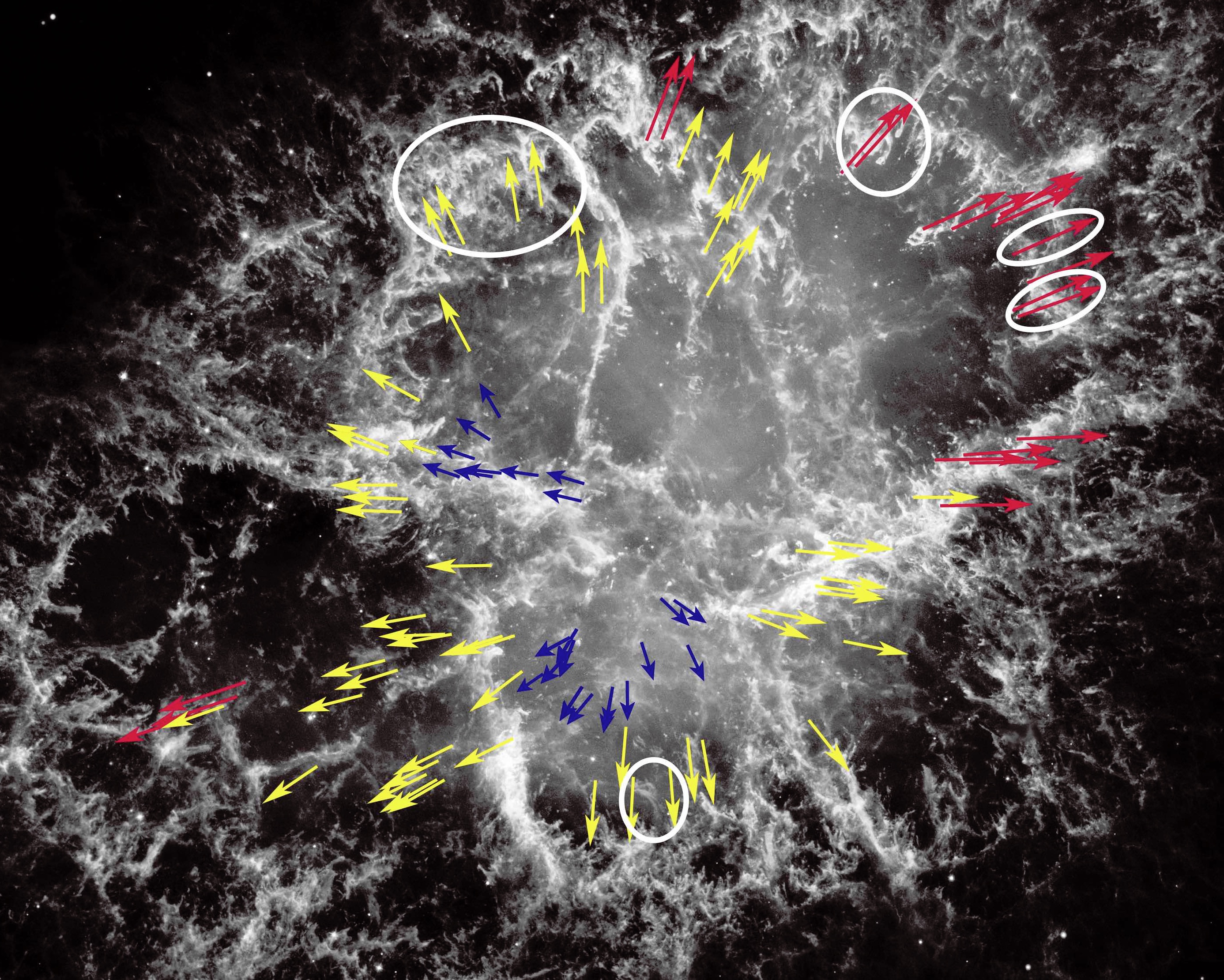}}
\caption{Transversal velocities are depicted with arrows, and grouped in three velocity intervals $<$ 400 (blue), 400--1000 (yellow), and $\geq $ 1000 (red) km s$^{-1}$ on the background of the HST image shown in Fig.~\ref{crab}. Encircled arrows mark irregular objects not included in Table A.} 
\label{fig:flow}
\end{figure}

\subsection{Velocity patterns}
\label{sec:flow}

From the proper motion measurements reported in Sect.~\ref{sec:obs} we obtain the position angle for the transversal velocity vector. The globules range in velocity from a minimum of 60~km~s$^{-1}$ to a maximum of 1600 km s$^{-1}$, significantly lower than the 2310 km s$^{-1}$ measured for the outer edge of the nebula by Bietenholz \& Nugent (\cite{bie15}). Fig.~\ref{fig:flow} shows directions of motion for all globules measured, where the velocities have been sorted into three velocity intervals: $<$ 400, 400--1000, and $\geq$ 1000 km s$^{-1}$. In this figure we have also included transversal velocities measured for some rather irregular dust features in the peripheries of the remnant, which are not listed in Table A.1, but are encircled in Fig.~\ref{fig:flow}. 

Obviously, the globules take part in the general expansion of the supernova remnant, and comparing with the transversal velocity vectors derived for emission filaments in Trimble (\cite{tri68}), for example, we conclude that the globules follow the same flow pattern as the emitting gas. The smallest velocities cluster close to the centre of the nebula, while globules with the largest velocities are located in the periphery. This is expected from a more or less spherically expanding constellation of globules. An asymmetry in the flow is also apparent, oriented along the major axis of the remnant. The proper motion of the pulsar is small compared to the transversal expansion velocities (Kaplan et al. \cite{kap08}; Wykoff \& Murray \cite{wyk77}). By tracing the velocity vectors backwards in time, we find an origin for the expanding system of globules within a radius of a few arcseconds from the divergent point of the remnant, according to Kaplan et al. (\cite{kap08}) to be  at 5$^{h}$~34$^{m}$~33$^{s}$  +22$\degr$~00$\arcmin$~48$\arcsec$ (J2000.0).    
 
Five globules, CrN 8, 12, 16, 30, and 31, are coincident in position with H$_{2}$ knots measured for radial velocity in Loh et al. (\cite{loh11}) for which we can derive the corresponding space velocities. Two objects, CrN 8 and 12, are in the outskirts of the remnant (Field 2 in Fig.~\ref{fields}). For these objects, the transversal velocities are large, but the radial velocities are small (+121 km s$^{-1}$ and +146 km s$^{-1}$), similar to the +140 km s$^{-1}$ measured for the ArH$^{+}$ emission in this region by Barlow et al. (\cite{bar13}). The space velocities for the two objects are large; approximately 1300 km s$^{-1}$. For globules CrN 16 and 31 in Fields 3 and 4, with radial velocities of +696 and -525 km s$^{-1}$, the resulting space velocities are smaller.

\section{Discussion}  
\label{sec:disc}

The mass of dust contained in the Crab Nebula globules and the distribution of sizes are in the same range as for the smallest globulettes residing in \ion{H}{ii} regions surrounding young clusters (e.g., Grenman \& Gahm \cite{gre14}), but larger than dusty knots found in certain planetary nebulae (O'Dell \& Handron \cite{ode96}). In contrast to these knots, and some of the globulettes, the Crab Nebula globules lack any signs of extended tails, simply because the globules do not interact with any central source of UV radiation. For the globules listed in Appendix~A, we find a mean density of dust of $2.5\cdot 10^{-21}$ g cm$^{-3}$, which is comparable to the corresponding values derived for globulettes in, for instance, the Rosette Nebula (Gahm et al.~\cite{gah07}). We note that typical masses of dust in the Crab Nebula globules are two orders of magnitude larger than the dusty clumps assumed by Ercolano et al. (\cite{erc07}) to match the energy distribution of SNR 1987A.  

The total mass of dust, $4.5\cdot 10^{-4}$ $M_\odot$ , as derived for all 92 globules, can be compared to estimates of the total mass of dust based on images of the infrared emission obtained by {\it Spitzer} and {\it Herschel}. This dust is spread over the entire nebula, but strongly concentrated to the filaments extending across the nebula (see Fig.~\ref{fig:positions}, left panel). Such estimates range from 0.02 -- 0.13 $M_\odot$  (Temim \& Dwek \cite{tem13}) to 0.18 -- 0.27 $M_\odot$ (Owen \& Barlow \cite{owe15}; Gomez et al. \cite{gom12}). Compared to these estimates, the globules contribute between 0.2  and 2 \% of the total mass. The globules are not resolved in the infrared images and might not have been properly sampled in the studies of the total dust content. Nevertheless, the dust in globules comprise only a small fraction of the total dust mass. The amount of dust contained in the filaments marked in the figures in Appendix~B is comparable to the total dust mass of the globules.

From our measurement of extinction in the V band and at 8140 \AA,\ we conclude that the extinction matches standard interstellar law. Since the globules are shooting out from the centre at high speed, one expects that they supply the interstellar medium with small grains with extinction properties similar to that of normal interstellar grains.

The origin of the dust grains present in supernova remnants has been subject to several studies. It has been proposed that dust is formed from material inside the stellar core shortly after explosion (e.g., Matsuura et al. \cite{mat15}, and references therein). Model calculations of such grain production have been presented in Kozasa et al. (\cite{koz09}), for example. At a later phase, the dust spreads into the expanding ejecta, where it can be further processed. In the Crab Nebula, the widespread dust is concentrated to the emission filaments as evidenced from {Herschel} observations. These filaments may be a consequence of Rayleigh-Taylor instabilities between the synchrotron nebula and the denser ejecta (Hester \cite{hes96}) indicating that the dust becomes concentrated along the filaments through the same process (Sankrit et al. \cite{san98}). Carlqvist (\cite{car15}) proposed that electromagnetic forces play a major role in the formation of certain finger-like filaments in the Crab Nebula, and one may consider it possible that plasma instabilities contribute to the formation of globules inside the warm filaments. Grain destruction in reverse shocks in supernovae ejecta can delete the population of small grains as modeled in Silvia et al. (\cite{sil12}), for example. Temim et al. (\cite{tem12}) pointed out that the clumping of gas and dust in the filaments may provide shielding against destruction. 

However, the majority of the globules are not located in any bright filaments, and we have no clear idea of how they have formed and survived. Model calculations of the convective layers in red supergiants have been presented by Chiavassa et al. (\cite{chi09}), for example, and Freytag (\cite{fre16}) predicts that a number of small convective surface elements (granules) cover the surface above the deeper part of the convection zone. We note that not only is the expected number of such granules of the order of one hundred, similar to the total number of globules we have listed, but also that their typical mass is comparable to the masses of globules corrected for a gas-to-dust ratio of 100. The agreement between these quantities may be a pure coincidence but might be worth considering in more detail. Could the globules in the Crab Nebula be remnants of gas cells in the outer layers of the progenitor, cells that are confined, compressed and ejected during the passage of the blast wave through the stellar surface and containing the seeds of grains that are further processed during the expansion?

\section{Conclusions}
\label{sec:conclude}

We have located 92 cold cloudlets seen in obscuration against the background of continuous synchrotron radiation in optical images of the Crab Nebula. These globules appear as dark, roundish spots in broad and medium band HST images. We present a list of positions, average radii, and orientations of elongated objects. From images taken between 1994 and 2014 we derived the proper motion and related transversal velocity of each globule. We estimate the amount of extinction in the V band for each object and with different assumptions on the amount of foreground emission. From these measures we obtained the individual masses of dust using two different methods. In addition, we estimated the mass of extended, obscuring filaments also present in the nebula. From extinction measurements at wavelengths 5470 \AA\ and 8140 \AA\ we acquire information on the shape of the extinction law for the globules.  

* Mean radii range from 400 to 2000 AU, and 40 \% of the globules are slightly elongated with a ratio between major and minor axes $>$ 1.5. The objects are too small to be distinguished in current infrared images obtained with {\it Spitzer} and {\it Herschel}.

* The globules are distributed over the entire nebula but less abundant in the region surrounding the central pulsar. Several globules are located at emission filaments where elongated objects are aligned with the filaments.

* The extinction law for the globules and filaments matches a normal interstellar extinction law. Hence, the globules will eventually feed the interstellar medium with grains of interstellar type.

* Derived masses of dust range from 1 to $60\cdot 10^{-6}$ $M_\odot$, with mean densities of $\sim$ $2.5\cdot 10^{-21}$ g cm$^{-3}$.

* The total mass contained in globules amounts to $4.5\cdot 10^{-4}$, a fraction of 2 \% at most of the total dust content of the nebula. Assuming a gas-to-dust content of 30, the corresponding total mass amounts to $1.4\cdot 10^{-2}$ $M_\odot$.

* All globules move outwards from the centre with transversal velocities of 60 to 1600 km s$^{-1}$ along with the general expansion of the remnant. For some objects we could also derive space velocities. 

The question about the origin of dust in supernova remnants remains open. Dust grains may form at an early stage of the explosion and be further processed and grow during the expansion of the ejecta. Rayleigh-Taylor instabilities acting on the ejecta may form filaments in which dust production and growth are stimulated, and we point out that electro-magnetic forces may play a role in confining globules inside filaments. However, most globules are not residing in filaments, and we speculate that they could be products of cell-like blobs in the atmosphere of the progenitor, cells that are collapsing to form globules during the passage of the blast wave during the explosion, possibilities which can be tested further with hydrodynamic calculations. The Crab Nebula globules can be resolved from infrared imaging with the coming James Webb Space Telescope, which would provide clues to whether the dust is warm or cold.

\begin{acknowledgements}

We thank Peter Lundqvist and Bernd Freytag for valuable comments. The suggestions from the referee T. Haworth were most helpful in improving our presentation. This work was supported by the Magnus Bergvall Foundation.

\end{acknowledgements}

\begin{appendix}
\label{Appendix A}

\section{Properties of the globules in the Crab Nebula.}

\begin{table*}
\centering
\caption{Symbols are described in Section~\ref{sec:obs}.}
\begin{tabular} {lccccccccccl}
 \hline\hline
     \noalign{\smallskip}
CrN & Field & R.A. & Dec. & $\bar r$ & $\bar R $ & P.A. &   $\mu_{\alpha}$ & $\mu_{\delta}$ &V$_{t}$ & $M_{d}$ & Remarks\\
 && J(2000.0) &J(2000.0) & ($\arcsec$) &  (AU) & (degr.)& ($\arcsec$/yr)&($\arcsec$/yr) & (km/s) & ($M_\odot$$\cdot 10^{-6}$) &\\       \noalign{\smallskip}   
\hline 
      \noalign{\smallskip}
       
1 &2&   5:34:25.77 &+22:01:40.1 &       0.5&    1000&      41&          -0.163  &       +0.064&1660 &        10 & \\
2 &1    &5:34:25.86&+22:01:03.6 &       0.3&     600&   -16     &       -0.133  &       +0.016  &1270&   3       &   \\
3 &     2&5:34:26.25&   +22:01:56.1     &       0.3&     600&           &               -0.145  &       +0.077  & 1522&    3      & \\
4 &2    &5:34:26.31&+22:01:54.9 &       0.2&     400&           &       -0.134  &       +0.093  & 1546&    1      &       \\
5&      2&5:34:26.34&   +22:01:54.3     &       0.2&     400&   -33     &               -0.143  &       +0.078  & 1544&    2      &       \\
6&1&    5:34:26.84&     +22:00:58.3     &       0.2&     400&           &               -0.113  &       +0.009  & 1075&    2      &       \\
7&      1&5:34:26.89&   +22:01:00.4     &       0.3&     600&   -42     &               -0.114  &       +0.016  & 1091&   3       &                      \\
8&2     &5:34:27.06&    +22:01:52.5     &       0.5&    1000&           &               -0.114  &       +0.071  & 1273&   12      &Knot 52$^{1}$, 4E$^{2}$ \\
9&      1&5:34:27.20&   +22:00:58.5     &       0.2&     400&           &               -0.113  &       +0.009  & 1075&   1   &C to 10                            \\
10&1&   5:34:27.21&     +22:00:57.7     &       0.4&     800&           &               -0.118  &       +0.004  & 1119&   6       & C to 9, Irr        \\
11&     1&5:34:27.25&   +22:00:49.7     &       0.6&    1200&           &               -0.108  &       +0.022  & 1045&   20  &1E$^{2}$, Irr               \\
12&     2&5:34:27.58&   +22:01:52.0     &       0.6&    1200&           &               -0.121  &       +0.073  & 1340&   16   &Knot 51$^{1}$, 4D$^{2}$    \\
13&1&5:34:27.67&        +22:00:52.2     &       0.2&     400&           &               -0.099  &       +0.008  & 942 & 2      &                                 \\
14&     3&5:34:28.63&   +22:00:19.9     &       0.3&     600&   -34     &               -0.081  &       -0.024  & 801 & 4 &                                \\
15&     3&5:34:29.01&   +22:00:42.1     &       0.2&     400&           &               -0.076  &       -0.001  & 721&    2  &                             \\
16&     3&5:34:29.32&   +22:00:30.1     &       1.0&    2000&           &               -0.068  &       -0.011  & 653&  58 & Irr, Knot 53$^{1}$, 1A$^{2}$\\
17&     3&5:34:29.37&   +22:00:27.9     &       0.3&     600&           &               -0.068  &       -0.016  & 662 & 3&                                \\
18&     3&5:34:29.49&   +22:00:30.3     &       0.6&    1200&   -44     &               -0.077  &       -0.012  & 739&    15&                                        \\
19&     5&5:34:29.53&   +22:00:00.5     &       0.2&     400&   22      &               -0.048  &       -0.051  & 664&    2&                      \\
20&     3&5:34:29.59&   +22:00:28.7     &       0.2&     400&           &               -0.061  &       -0.012  & 589 & 2      &                                \\
21&     3&5:34:29.59&   +22:00:40.5     &       0.2&     400&   -4      &               -0.061  &       -0.004  & 580&    3       &           \\
22&     3&5:34:29.66&   +22:00:29.2     &       0.3&     600&    31     &               -0.060  &       -0.015  & 586&    4       &                                \\
23&     3&5:34:30.24&   +22:00:24.9     &       0.2&     400&    3      &               -0.052  &       -0.010  & 502&    2       &                                  \\
24&     3&5:34:30.44&   +22:00:25.0     &       0.2&     400&           &               -0.045  &       -0.023  & 479&    2       &                                \\
25&     4&5:34:30.71&   +22:01:57.9     &       0.3&     600&   40      &               -0.041  &       +0.081  & 861&    4       &                      \\
26&     4&5:34:30.75&   +22:01:56.0     &       0.2&     400&           &               -0.041  &       +0.092  & 955 &  3&                               \\
27&     6&5:34:30.79&   +22:01:40.3     &       0.4&     800&   15      &               -0.039  &       +0.050  & 601&6   &        4B$^{2}$              \\
28&     5&5:34:31.06&   +21:59:53.9     &       0.2&     400&           &               -0.023  &       -0.057  & 583&2   &                                 \\
29&     6&5:34:31.08&   +22:01:37.4     &       0.3&     600&           &               -0.042  &       +0.063  & 718&2   &                                    \\
30&     6&5:34:31.15&   +22:01:47.2     &       0.7&    1400&           &               -0.046  &       +0.073  & 818&16  & Knot 45$^{1}$, 4C$^{2}$  \\
31&     4&5:34:31.19&   +22:02:01.0     &       0.6&    1200&   -21     &               -0.035  &       +0.086  & 880&17  & Knot 46$^{1}$, 4A$^{2}$     \\
32&     5&5:34:31.20&   +22:00:11.7     &       0.2&     400&   21      &               -0.025  &       -0.032  & 385&    2&                      \\
33&     5&5:34:31.44&   +21:59:56.0     &       0.4&     800&   44      &               -0.016  &       -0.054  & 534&6   &                                      \\
34&     3&5:34:31.63&   +22:00:26.6     &       0.3&     600&           &               -0.018  &       -0.015  & 222&    4&                                \\
35&     4&5:34:31.64&   +22:02:07.0     &       0.3&     600&-7 &               -0.042  &       +0.089  & 933 &4       &                                \\
36&     4&5:34:31.67&   +22:02:14.3     &       0.3&     400&   -16     &               -0.027  &       +0.104  & 1019&3  &                                \\
37&4&5:34:31.72&        +22:02:13.6     &       0.2&     400&   40      &               -0.025  &       +0.104  & 1014&2  &                                   \\
38&     3&5:34:31.76&   +22:00:28.3     &       0.4&     600&           &               -0.012  &       -0.015  & 182 &5       &                                 \\
39&     3&5:34:32.20&   +22:00:18.0     &       0.2&     400&           &               -0.006  &       -0.021  & 207 &   1&                             \\
40&     5&5:34:32.20&   +21:59:59.4     &       0.2&     400&   -41     &               +0.009  &       -0.050  & 482&1   &                                 \\
41&     5&5:34:32.36&   +22:00:06.2     &       0.3&     600&           &               -0.004  &       -0.032  & 306&3        &                                 \\
42&     5&5:34:32.62&   +22:00:06.6     &       0.2&     400&   -1      &               +0.005  &       -0.034  & 326&2   &        2C$^{2}$           \\
43&5&   5:34:32.62&     +22:00:05.5     &       0.3&     600&           &               +0.005  &       -0.036  & 345&4   &       2C$^{2}$                \\
44&6&   5:34:32.85&     +22:01:35.6     &       0.2&     400&           &               +0.002  &       +0.053  & 503&2   &                                 \\
45&5&   5:34:32.90&     +21:59:45.0     &       0.2&     400&   34      &               +0.013  &       -0.059  & 573&1&                      \\
46&5&   5:34:33.04&     +22:00:02.4     &       0.6&    1200&           &                       +0.021  &       -0.029  & 339&11  &       2D$^{2}$, Irr   \\
47&     5&5:34:33.15&   +22:00:05.9     &       0.3&     600 &  39      &               +0.013  &       -0.035  & 354&4&                                  \\
48&6&   5:34:33.19&     +22:01:33.1     &       0.4&     800&           &                       +0.005  &       +0.060  & 571&6   &       Irr   \\
49&     7&5:34:33.18&   +22:00:54.4     &       0.3&     600&   -35     &               +0.018  &       +0.003  & 173&4&                 \\
50&     7&5:34:33.18&   +22:00:49.5     &       0.2&     400&           &                       +0.007  &       +0.002  &  69&    2       &               \\
51&6&   5:34:33.25&     +22:01:42.6     &       0.2&     400&           &                  +0.012     &   +0.069 &664  &  1      &                               \\
52&8&   5:34:33.32&     +22:00:20.5     &       0.4&     800&   3       &                       +0.016  &       -0.021  & 250&6   &C to 53, 2E$^{2}$   \\
53&8&5:34:33.38&        +22:00:18.3     &       0.4&     800&   27      &               +0.017  &       -0.032  & 344&7&  C to 52, 2E$^{2}$                    \\
54&8&   5:34:33.44&     +22:00:20.2     &       0.3&     600&   -36     &               +0.019  &       -0.016  & 236&4   &                                \\
55&     7&5:34:33.47&   +22:00:53.0     &       0.2&     400&           &               +0.008  &       +0.007  & 101&1   &                        \\
56&     8&5:34:33.50&   +22:00:19.4     &       0.3&     600&           &               +0.026  &       -0.018  & 300&3   &                              \\
57&8&5:34:33.52&        +22:00:16.5     &       0.3&     600&           &               +0.024  &       -0.022  & 309 &4       &                                 \\
58&     8&5:34:33.66&   +22:00:15.3     &       0.2&     400&           &               +0.029  &       -0.023  & 351& 2       &                                 \\
59&     7&5:34:33.96&   +22:00:54.2     &       0.2&     400&           &           +0.021  &  +0.002 &200 & 2    &                                \\
60&7&   5:34:34.12&     +22:00:54.3     &       0.2&     400&           &               +0.034  &       +0.005  & 326&2&                 \\

     \noalign{\smallskip}
\hline

\end{tabular}
\end{table*}

\begin{table*}
\centering
\caption{Symbols are described in Section~\ref{sec:obs}.}
\begin{tabular} {lccccccccccl}
 \hline\hline
     \noalign{\smallskip}
CrN & Field & R.A. & Dec. & $\bar r $ & $\bar R $ & P.A. &  $\mu_{\alpha}$ & $\mu_{\delta}$ & V$_{t}$ & $M_{d}$ & Remarks\\
 && J(2000.0) &J(2000.0) & ($\arcsec$) &  (AU) & (degr.)& ($\arcsec$/yr)&($\arcsec$/yr) & (km/s) & ($M_\odot$$\cdot 10^{-6}$) &\\\noalign{\smallskip}   
\hline 
     \noalign{\smallskip}

61&9&   5:34:34.17&     +22:01:12.9     &       0.2&     400&           &               +0.024  &       +0.024  & 322 & 1      &                 \\
62&     10&5:34:34.29&  +21:59:55.6     &       0.2&     400&           &               +0.039  &       -0.044  &557  & 1    &               \\
63&     8&5:34:34.31&   +22:00:11.0     &       0.3&     600&           &               +0.034  &       -0.027  &412  & 3    &                      \\
64&7&   5:34:34.66&     +22:00:35.3     &       0.4&     800&           &               +0.043  &       -0.002  &408   &6    &           \\
65&9&   5:34:34.69&     +22:01:02.6     &       0.4&     800&   -3      &               +0.039  &       +0.015  & 396&8&          3E$^{2}$     \\
66&     7&5:34:34.69&   +22:00:56.3     &       0.3&     600&   -31     &               +0.024  &       +0.005  & 232&5   &              \\
67&8&   5:34:34.76&     +22:00:19.8     &       0.2&     400&           &               +0.044  &       -0.018  &451  &  2&     C to 69            \\
68&9&   5:34:34.83&     +22:01:26.0     &       0.2&     400&   24      &               +0.030  &       +0.038  &459&1  &                                     \\
69&8&   5:34:34.84&     +22:00:19.1     &       0.5&    1000&   33      &               +0.043  &       -0.016  &435  &  11&         C to 67, 2F$^{2}$     \\
70&7&5:34:34.85&   +22:00:54.2  &       0.3&     600&           &                +0.032  &       +0.008          & 313 &  3       &                    \\
71&10&5:34:35.18&       +21:59:53.3     &       0.3&     600&           &               +0.055  &       -0.046           & 680 & 4&                        \\
72&7&5:34:35.40&        +22:00:58.9     &       0.4&     800&   30      &               +0.037  &       +0.017                 &  386 & 8      &       3G$^{2}$   \\
73&10&5:34:35.41&       +21:59:42.6     &       0.2&     400&           &               +0.059  &       -0.057          & 778 & 2 &                            \\
74&8&5:34:35.58&        +22:00:23.8     &       0.2&     400&   -7      &                 +0.059 &       -0.007          &  563 &3  &             C to 75    \\
75&8&5:34:35.60&        +22:00:23.1     &       0.3&     600&   -19     &               +0.076  &       -0.005                 & 722  & 4 &    C to 74        \\
76&10&5:34:35.65&       +21:59:49.4     &       0.2&     400&           &               +0.068  &       -0.049          & 795 &    2      &                  \\
77&8&5:34:35.76&        +22:00:26.6     &       0.2&     400&           &               +0.067  &       -0.016          & 653 &2  &                         \\
78&9&5:34:35.87&        +22:01:12.8     &       0.3&     600&   43      &               +0.047  &       +0.031          & 534 & 4 &                  \\
79&10&5:34:35.92&       +21:59:47.4     &       0.2&     400&           &                 +0.064  &       -0.054                 & 794 & 2 &                     \\
80&7&5:34:36.02&        +22:00:50.5     &       0.3&     600&           &               +0.063  &       +0.004            &  598 &       5       &       C to 81\\ 
81&7&5:34:36.09&        +22:00:52.1     &       0.4&     800&           &               +0.058  &       +0.007          & 554 &7  &       C to 80   \\
82&8&5:34:36.12&        +22:00:11.4     &       0.2&     400&                   &         +0.077          &   -0.022              &  759&  2 &                             \\
83&7&5:34:36.15&        +22:00:46.8     &       0.3&     600&           &               +0.072  &       +0.004          &  684&  4  &                         \\
84&9&5:34:36.44&        +22:01:01.9     &       0.2&     400&           &               +0.058  &       +0.020          &  582 & 2        &       C to 85\\
85&9&5:34:36.45&        +22:01:02.5     &       0.2&     400&   -5      &               +0.060  &       +0.019          &  597&   2       & C to 84    \\
86&8&5:34:36.52&        +22:00:12.8     &       0.2&     400&           &               +0.084  &       -0.019          &   816&  2   &                       \\
87&8&5:34:36.95&        +22:00:07.8     &       0.3&     600&           &               +0.097  &       -0.020          &  939&   4       &            \\
88&10&5:34:37.57&       +21:59:49.3     &       0.5&    1000&   -38     &               +0.086  &       -0.060          &  994 & 9        &                   \\
89&11&5:34:38.73&       +22:00:08.0     &       0.2&     400&           &               +0.117  &       -0.041          &  1175 & 1 &                       \\
90&11&5:34:38.82&       +22:00:05.4     &       0.2&     400&           &               +0.120  &       -0.034          &  1182  & 2      &                  \\
91&11&5:34:39.06&       +22:00:03.1     &       0.2&     400&   -5      &               +0.091  &       -0.044          &  958 &  2         &                \\
92&11&5:34:39.38&       +22:00:02.1     &       0.2&     400&           &               +0.100  &       -0.049          &1056 &       1       &                 \\              

     \noalign{\smallskip}
\hline
\end{tabular}
\tablefoot{
\tablefoottext{1}{from Loh et al. (\cite{loh11}),}
\tablefoottext{2}{from Fesen \& Blair (\cite{fes90})}
}
 
\end{table*} 

\end{appendix}

\begin{appendix}
\label{Appendix B}

\section{Images of all fields listed in Table~\ref{tab:regions}. }

\begin{figure*}[t] 
\centering
\resizebox{16cm}{!}{\includegraphics[angle=00]{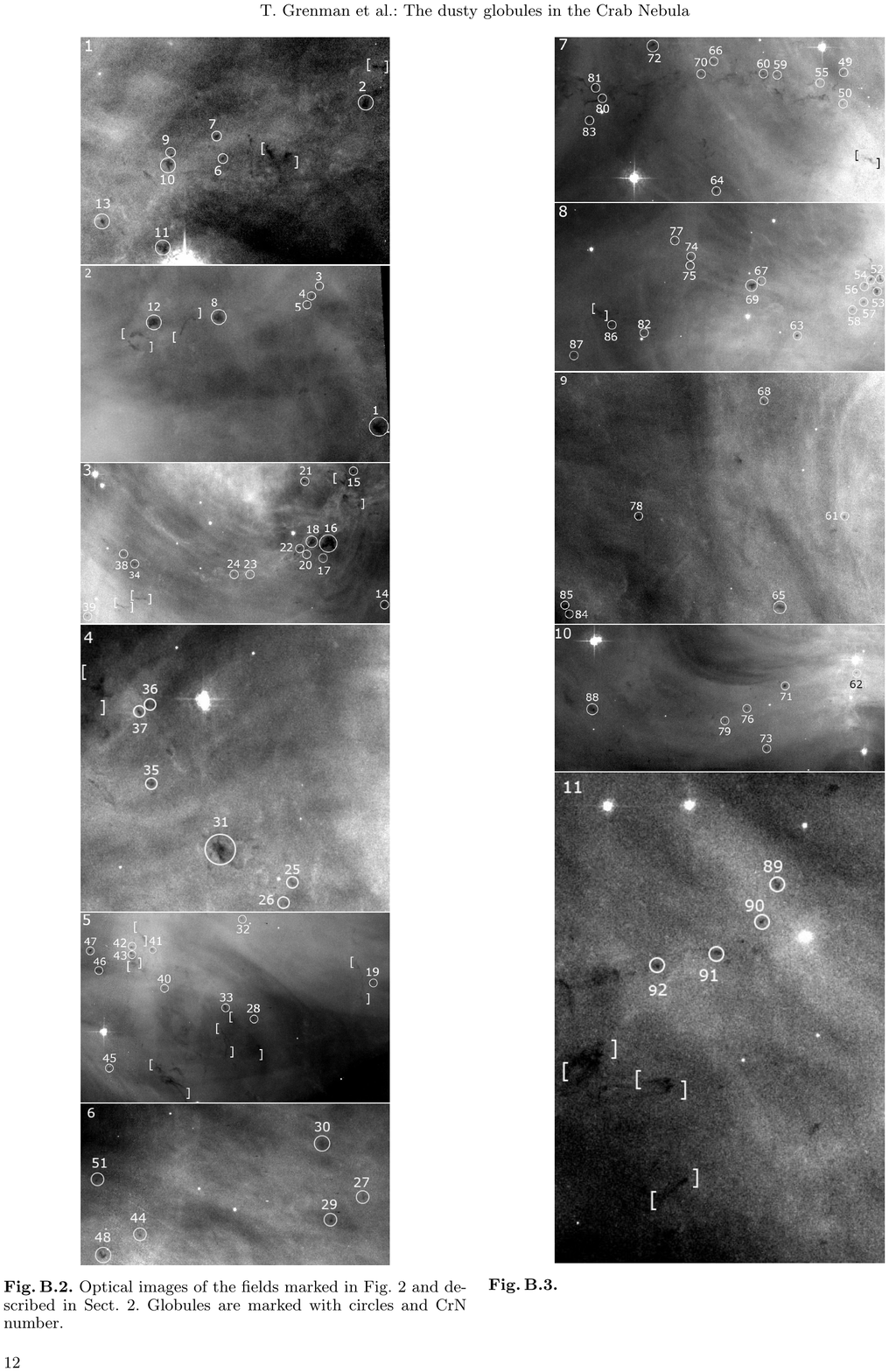}}
\caption{Optical images of the fields marked in Fig.~\ref{fields} and described in Sect.~\ref{sec:obs}. Globules are marked with circles and CrN number. } 
\label{map1}
\end{figure*}

\end{appendix}

\end{document}